\documentclass[a4paper,twocolumn,english,showpacs]{revtex4}
\usepackage[T1]{fontenc}
\usepackage[latin1]{inputenc}
\usepackage{amsmath}
\usepackage{graphicx}
\usepackage{amssymb}

\makeatletter
\usepackage{babel}
\makeatother
\begin{document}

\title{Fractal Conductance Fluctuations of Classical Origin}

\author{H. Hennig$^{1,2}$, R. Fleischmann$^{1,2}$, L. Hufnagel$^{3}$,
T. Geisel$^{1,2}$}

\address{$^{1}$Max Planck Institute for Dynamics and Self-Organization, 37073
Göttingen, Germany}

\address{$^{2}$Department of Physics, University of Göttingen, Germany}

\address{$^{3}$Kavli Institute for Theoretical Physics, University of California,
Santa Barbara, USA }

\date{\today}

\begin{abstract}
In mesoscopic systems conductance fluctuations are a sensitive probe
of electron dynamics and chaotic phenomena. We show that the conductance
of a purely classical chaotic system with either fully chaotic or
mixed phase space generically exhibits fractal conductance fluctuations
unrelated to quantum interference. This might explain the unexpected
dependence of the fractal dimension of the conductance curves on the
(quantum) phase breaking length observed in experiments on semiconductor
quantum dots.
\end{abstract}


\maketitle
A prominent feature of transport in mesoscopic systems is that the
conductance as a function of an external parameter (e.g. a gate voltage
or a magnetic field) shows reproducible fluctuations caused by quantum
interference \cite{Marcus_PRL_92}. A prediction from semiclassical
theory that inspired a number of both theoretical and experimental
works in the fields of mesoscopic systems and quantum chaos was that
in chaotic systems with a mixed phase space these fluctuations would
result in fractal conductance curves \cite{Ketzmerick_PRB_96,hufnagel_01_epl}.
Such fractal conductance fluctuations (FCFs) have since been confirmed
in gold nanowires and in mesoscopic semiconductor billiards in various
experiments \cite{Hegger_96_PRL,Ketzmerick_PRL_98,Micolich_98_JPCM,Ochiai_98_SeScTech,Takagaki_00_PRB}.
In addition FCFs have more recently been predicted to occur in strongly
dynamically localized \cite{guarneri_01_PRE} and in diffusive systems
\cite{pinheiro_06_BJP}. Due to the quantum nature of the FCFs it
came as a surprise when recent experiments indicated that decoherence
does not destroy the fractal nature of the conductance curve but only
changes its fractal dimension \cite{Micolich_01_PRL,Micolich_02_PhysE}.
In the present letter we show, that the conductance of purely classical
(i.e.~incoherent) low-dimensional Hamiltonian systems very fundamentally
exhibits fractal fluctuations, as long as transport is at least partially
conducted by chaotic dynamics. Thus mixed phase space systems and
fully chaotic systems alike generally show FCFs with a fractal dimension
that is determined analytically. We show that it is governed by fundamental
properties of chaotic dynamics. 

In a disordered mesoscopic conductor -- which is smaller than the
phase coherence length of the charge carriers but large compared to
the average impurity spacing -- the transmission is the result of
the interference of many different, multiply-scattered and complicated
paths through the system. As these paths are typically very long compared
to the wave length of the charge carriers, the accumulated phase along
a path changes basically randomly when an external parameter such
as the energy or the magnetic field is varied. This results in a random
interference pattern, i.e. reproducible fluctuations in the conductance
of a universal magnitude on the order of $2e²/h$, the so called \emph{universal
conductance fluctuations} (UCFs). \emph{}For a review see \cite{book_imry}
or \cite{book_ferry}. The role of disorder in providing a distribution
of random phases can as well be taken by chaos. Thus ballistic mesoscopic
cavities like quantum dots in high mobility two-dimensional electron
gases that form c\emph{haotic billiards} show the same universal fluctuations
\cite{Jalabert_90_PRL,Marcus_PRL_92,Baranger_93_Chaos}. If the average
of \emph{}the \emph{phase gain} accumulated on the different paths
traversing the system exists, the conductance curves are smooth on
parameter scales that correspond to a change of the average phase
gain on the order of and smaller than the wave length of the carriers.
In systems with mixed phase space, where chaotic and regular motion
coexist, this phase gain, however, is typically algebraically distributed
and its average phase gain does not exist (neglecting the finiteness
of the coherence length and assuming the semiclassical limit $\hbar_{eff}\rightarrow0$
; the role of the finite $\hbar_{eff}$ is discussed in \cite{hufnagel_01_epl}).
Therefore, as shown in \cite{Ketzmerick_PRB_96}, the conductance
curve of such a system fluctuates on all parameter scales and forms
a fractal. The fractal dimension $D$ is connected to the exponent
$\gamma$ of the algebraic distribution of phase gains by $D=2-\frac{\gamma}{2}.$

Experiments on quantum dots that study the dependence of the conductance
fluctuations on several system parameters like size and temperature
seem to partly contradict the semiclassical theory of fractal scaling
\cite{Micolich_01_PRL,Micolich_02_PhysE}. Namely it was found that
with decreasing coherence length the scaling region over which the
fractal was observed did not shrink -- as would be expected from the
semiclassical arguments --, but that the fractal dimension changed.
An implicit assumption of the semiclassical theory is that the classical
dynamics remains unchanged as the external parameter is varied and
thus only phase changes are relevant. In most experimental setups,
however, the classical phase space changes with variation of the control
parameter. In this article we show that the classical chaotic dynamics
itself already leads to fractal conductance curves! Moreover, from
this follows that even on very small parameter scales the fluctuations
due to changes in the classical dynamics are important. In general
the conductance curve is a superposition of two fractals: one originating
in interference which is suppressed by decoherence to reveal the fractal
fluctuations reflecting the changes in the classical phase space structure.
In addition, we predict that FCFs are not only observable in systems
with a mixed phase space but in purely chaotic systems. 

\begin{figure}[bh]
\includegraphics[width=8.5cm,keepaspectratio]{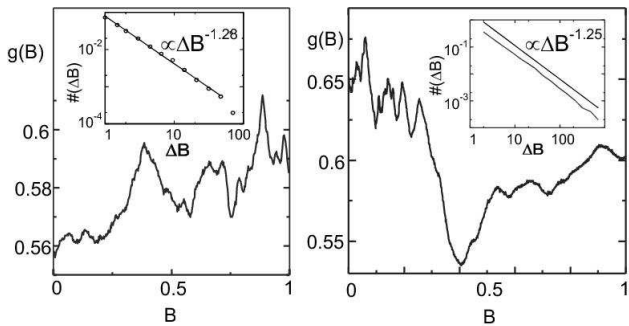}

\caption{Classical Conductance $g(B)$ through a stadium (left, geometry as
in ref.~\cite{Ketzmerick_PRL_98}) and square billiard (right, geometry
as in ref.~\cite{Micolich_01_PRL}) versus magnetic field $B$. Both
fluctuating conductance curves are fractals (see insets and text).
Their respective dimensions are $D\approx1.28$ for the stadium and
$D\approx1.25$ for the square billiard. The fractal dimensions are
in good agreement with experimental measurements \cite{Ketzmerick_PRL_98,Micolich_01_PRL}.
\label{cap:billiards}}
\end{figure}
As a starting point of our investigations and to connect it to the
experiments we numerically study the classical conductance through
a rectangle (hard-wall) and a stadium billiard (soft-wall) as a function
of a magnetic field as shown in Fig.~\ref{cap:billiards}. (Throughout
this article, we will study the transmission, which, in accordance
with the Landauer theory of conductance, is proportional to the conductance,
see e.g.~\cite{book_datta}.) Note that not only the phase space
of the stadium but also of the rectangle billiard is mixed in the
presence of a perpendicular magnetic field. In both cases, a modified
version \cite{note_variation,book_meakin} of the box-counting analysis
clearly reveals \textit{the fractal} nature of the conductance curves.
As the simulation is purely classical, the fractal scaling cannot
be caused by interference effects. So what is the underlying mechanism
for the fractality of the conductance curve and how can we understand
its dimension?

\begin{figure}[t]
\includegraphics[width=8cm,keepaspectratio]{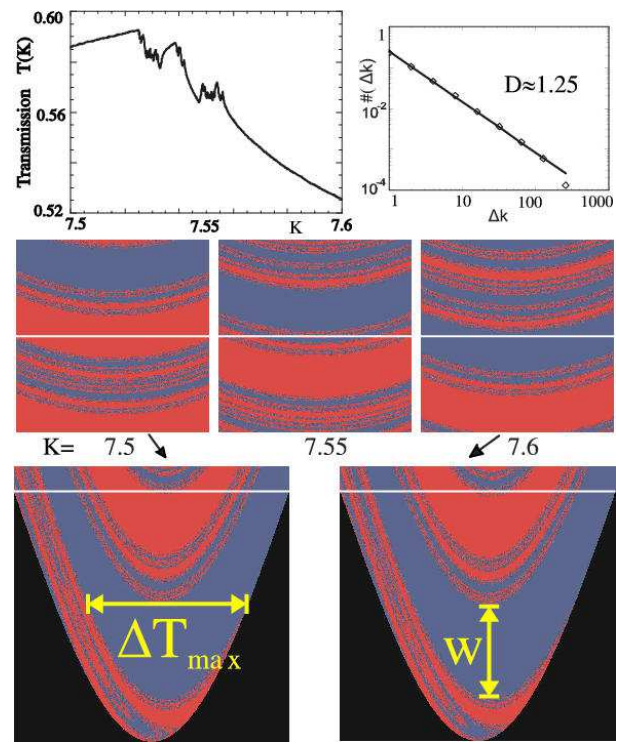}

\caption{How lobes translate into fluctuations: In the lower row the entryset
of the standard map with absorbing boundary conditions at $\pm3\pi$
for $K=7.5$ and $K=7.6$ resp.~can be seen. The three pictures in
the center row show the magnification of the central sections of the
entryset for three different values of $K=7.5$, $7.55$ and $7.6$.
The transmission $T(K)$ for $K=7.5\ldots7.6$ \cite{note_accmodes}
is shown in the top left picture. Note that a small change in $K$
shifts the lobes vertically, but conserves the overall phase space
structure, and that the largest fluctuations are caused by intersection
with the apex of lobes. Starting from $K=7.5$, a large transmission
lobe is cut by the horizontal line (see text), i.e.~the transmission
increases with $K$. In the same way, e.g.~the fluctuations of $T(K)$
near $K=7.55$ can be understood. The box-counting analysis reveals
a fractal structure (top right). \label{cap:lobes2fluc}}
\end{figure}
To study this mechanism in detail we will, because of its numerical
and conceptual advantages, analyze the transport in Chirikov's standard
map \cite{chirikov_79_PR,Fishman_82_PRL,Altland_96_PRL}. This paradigmatic
system shows all the richness of Hamiltonian chaos. And since -- as
will become apparent below -- our theory relies only on very fundamental
properties of chaotic systems, it is a natural choice as a model system.
The standard map is defined by\begin{eqnarray*}
\theta^{\prime} & = & p+\theta\\
p^{\prime} & = & p+K\sin\theta^{\prime}\end{eqnarray*}
with momentum $p$ , angle $\theta$ and the 'nonlinearity parameter'
$K$, which drives the dynamics from fully integrable ($K=0$) to
fully chaotic ($K\gtrsim7$). In between the phase space is mixed.
The standard map can be seen as the Poincaré surface of a conservative
system of two degrees of freedom. As such the map can by viewed to
directly correspond to the Poincaré map at the boundary of a chaotic
ballistic cavity, connecting it conceptually with the experimental
system. We introduce absorbing boundary conditions (see e.g.~ref.~\cite{Jacquod_04_PRL}),
i.e.~when $p$ exceeds (drops below) a maximum (minimum) threshold
value, the particle is transmitted (reflected) and leaves the cavity.
As can be seen right from the definition of the standard map, the
envelope of the entryset (which is, the phase space projection of
the injection lead) is simply half a period of a sine function times
$K$.

A trajectory entering the system eventually contributes either to
the total transmission or reflection, and we mark the corresponding
point in the entryset by a color code (transmission: red, reflection:
blue). Chaotic dynamics, through its fundamental property of stretching
and folding in phase space, leads to a lobe structure (see Fig.~\ref{cap:lobes2fluc}
(bottom)), which is typical for chaotic systems and not special to
the standard map. The distribution of widths $w$ of lobes exhibits
a power law\begin{equation}
n(w)\propto w^{-\alpha}.\label{eq:powerlaw_width}\end{equation}
 The lobe structure is translated into transmission by summing up
the intersections of the transmission lobes along a horizontal line,
see Fig.~\ref{cap:lobes2fluc}. A lobe of thickness $w$ gives rise
to a maximum contribution $\Delta T\propto w^{\beta}$. Variation
of the external parameter $K$ leads to a fractal transmission curve
$T(K)$ with $D\approx1.25$. 

\begin{figure}[h]
\includegraphics[width=8cm]{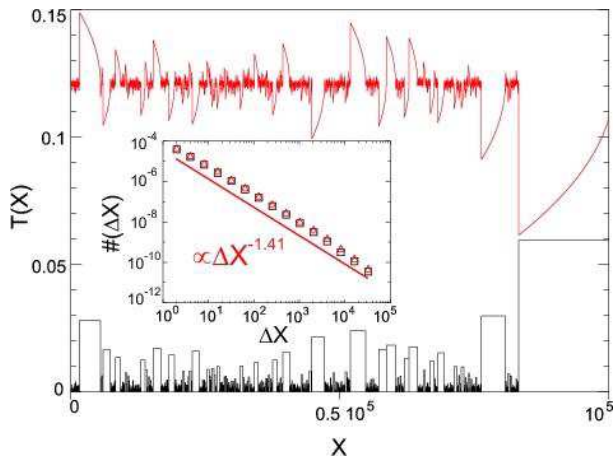}

\caption{Transmission $T(X)$ for lobes (red upper curve, shifted along the
y-axis for clarity) and stripes (black lower curve) for one and the
same random distribution with $\alpha=1.9$, $\beta=0.5$. The inset
shows the box-counting analysis for the upper (red triangles) and
lower transmission curve (black squares). The regression line is drawn
for the upper curve, whose fractal dimension is $1.41$. \label{cap:random_lobes}}
\end{figure}

How does the fractal dimension depend on the power law distribution
of lobe-widths and the curvature of the lobes? To this aim, we study
a random sequence of curve segments mimicking the intersection of
consecutive lobes of widths $w$, distributed algebraically with exponent
$\alpha$ and curved like $w^{\beta}$. We define $X_{i}:=\sum_{j=1}^{i}w_{j}$
and\[
T(X)=(-1)^{i}(X-X_{i})^{\beta}\quad\mbox{:}\quad X_{i}<X\le X_{i+1}.\]
 An example of this curve of {}``random lobes'' with $\alpha=1.9$
and $\beta=\frac{1}{2}$ is shown in Fig.~\ref{cap:random_lobes}
(top). The box-counting analysis clearly reveals a fractal structure.

We further simplify the problem by replacing the lobes by a sequence
of stripes of widths $x$ with power law distribution $n(x)\propto x^{\alpha}$.
Dispensing with the sign of the fluctuation, the transmission reads\[
T(X)=(X_{i+1}-X_{i})^{\beta}.\]
This yields histogrammatic transmission curves $T(X)$ like the bottom
curve of Fig.~\ref{cap:random_lobes}. As shown in the inset, the
fractal dimension of the resulting transmission curve remains unchanged
compared to the corresponding calculation with random lobes within
the precision of the box-counting analysis. Thus, the fractal dimension
of the curve does not change noticeably when considering stripes instead
of lobes and also when neglecting the sign of each contribution, confirming
the intuition, that the fractal dimension depends only on the relative
scaling, i.e.~$\alpha$ and $\beta$, but not on the detailed form
of the curve sections. 

\begin{figure}
\includegraphics[width=8cm]{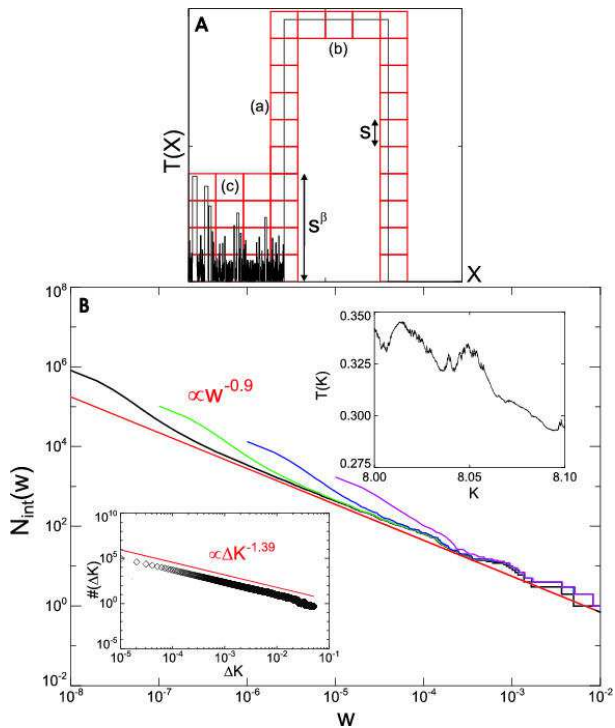}

\caption{\noindent \textbf{A.} Schematic transmission according to Fig.~\ref{cap:random_lobes}
(bottom), covered with boxes of size $s$. There are three contributions
marked (a-c). \textbf{B.} Total number $N_{int}(w)=\int_{w}^{\infty}n(w^{\prime})dw^{\prime}$
of lobes (for the open standard map with $|p|<4\pi$) of width larger
than $w$ on a double logarithmic scale. The four curves show estimates
for increasing resolution $w_{min}=10^{-5}\ldots10^{-8}$. The curves
clearly approach a power law corresponding to $n(w)\propto w^{-1.9}$.
The insets show the transmission curve $T(K)$ for values $K=8.0\ldots8.1$
calculated from $2\times10^{13}$ trajectories and its box-counting
dimension. \label{cap:comp}}
\end{figure}

For these curves like the bottom one of Fig.~\ref{cap:random_lobes}
with $\alpha-\beta>1$, we can give an analytical expression for the
fractal dimension and then estimate the fractal dimension of the transmission
curve in the standard map. We apply the box-counting method, which
we therefore review shortly (see e.g.~\cite{book_meakin} for a more
detailed introduction). In this approach the fractal curve lying in
a $n-$dimensional space is covered by a $n-$dimensional grid. Let
the grid consist of boxes of length scale $s$. The box-counting dimension
is then given by\begin{equation}
D=-\lim_{s\to0}\frac{\log N(s)}{log(s)},\label{eq:Dbox}\end{equation}
where $N(s)$ is the number of non-empty boxes. For our problem, we
divide $N(s)$ into three contributions $N(s)=n_{a}+n_{b}+n_{c}$,
as schematically drawn in Fig.~\ref{cap:comp}(A). The number $n_{a}$
of vertically placed boxes (see mark (a)) covering contributions from
stripes of widths $x>s$, reads\begin{equation}
n_{a}(s)\propto\frac{1}{s}\int_{s}^{\infty}p(x)x^{\beta}dx\propto s^{-(\alpha-\beta)}.\label{eq:n4}\end{equation}
Secondly, the number $n_{b}$ of horizontally placed boxes covering
horizontal contributions of stripes of widths larger than $s$, see
Fig.~\ref{cap:comp}A(b), is given by\begin{equation}
n_{b}(s)=\frac{1}{s}\int_{s}^{\infty}p(x)xdx<\frac{1}{s}\int_{0}^{\infty}p(x)xdx.\label{eq:nb}\end{equation}
Hence $n_{b}$ scales like $s^{-1}$ and can be neglected in comparison
to $n_{a}$ because of $\alpha-\beta>1$. Finally, we determine an
upper estimate for the number $n_{c}$ of vertically placed boxes
covering the contribution from stripes of widths $x\le s$. The total
length of these widths is $L(s)=\int_{0}^{s}p(x)xdx,$ therefor $L(s)/s$
boxes are needed to cover the length. Inflating all \emph{heights}
of the stripes $x\le s$ to the maximum possible size $s^{\beta}$,
see Fig.~\ref{cap:comp}A(c), we find\begin{equation}
n_{c}(s)<\frac{L(s)}{s}\,\frac{s^{\beta}}{s}\propto s^{-\alpha+\beta}.\label{eq:nc}\end{equation}
For $s\!\ll\!1$ thus the dominant terms is $n_{a}(s)$. With Eq.~(\ref{eq:Dbox}),
$N(s)$ gives rise to the box-counting dimension \cite{note_good_agreement}\begin{equation}
D=-\lim_{s\to0}\frac{\log s^{-\alpha+\beta}}{\log s}=\alpha-\beta.\label{eq:D_alpha_beta}\end{equation}

To connect the analytical result with the calculations of the transmission
of the open standard map, we numerically estimate the distribution
of lobe-widths in the entryset for $K=8$, finding $\alpha\approx1.9$,
as shown in Fig.~\ref{cap:comp}B. Together with $\beta=\frac{1}{2}$,
corresponding to first order Taylor expansion of the cosine function,
Eq.~(\ref{eq:D_alpha_beta}) predicts a fractal dimension $D\approx1.4$.
Direct analysis of the transmission curve (see insets of Fig.~\ref{cap:comp}B)
yields a fractal dimension $D\approx1.39$, in good agreement with
the expected value.

How can a power law distribution of lobe widths emerge in a fully
chaotic open system? One might rather expect to find an exponential
distribution of lobes in a fully chaotic system. To see why the distribution
is algebraic, however, let us examine the simplest case of an open
chaotic area preserving map the dynamics of which is governed by a
single, positive homogeneous Lyapunov exponent $\lambda$. In each
iteration phase space structures are stretched in one direction by
$\exp(\lambda)$, shrunk by $\exp(-\lambda)$ in the other and then
folded back. The entryset of the open system is thus stretched into
lobes of decaying width $w(t_{i})\propto\exp(-\lambda t_{i})$. The
phase space volume flux out of the system decays exponentially as
it is typical for a fully chaotic phase space, i.e. $\Gamma(t_{i})\propto\exp(-t_{i}/\tau),$
with (mean) dwelltime $\tau$. The area $\Gamma(t_{i})\Delta t$ is
the fraction of the exitset that leaves the system at time $t_{i}$.
With $t_{i}(w)\propto-\ln(w)/\lambda$ the number of lobes of width
$w$ in the exitset is \cite{note_entryset}\[
\# w=\frac{\Gamma(t_{i}(w))\Delta t}{w}\propto\frac{1}{w}\exp(\frac{\ln(w)}{\lambda\tau})=w^{\frac{1}{\lambda\tau}-1}.\]
 This suggests that the power law distribution of lobe widths is a
generic property even for fully chaotic systems. A quantitative expression
for the exponent, however, is not as easy to derive, as e.g.~the
Lyapunov exponent for the standard map is not homogeneous.

In conclusion, we have shown that transport through chaotic systems
due to the typical lobe structure of the phase space in general produces
fractal conductance curves, where the fractal dimension reflects the
distribution of lobes in the exit- /entryset. In contrast to the semiclassical
effect the size of the fluctuations is not universal but depends on
specific system parameters. Due to the fractal nature of the classical
conductance, however, there is no parameter scale that separates coherent
and incoherent fluctuations.

\bibliographystyle{apsrev}
\bibliography{fcf}

\begin{thebibliography}{26}
\bibitem[{not({\natexlab{a}})}]{note_variation}
\bibinfo{note}{{Throughout this article, in order to determine the fractal
  dimension of a given 2D curve $T(\Delta k)$, we used the so called variation
  method, i.e. we calculated $\# (\Delta k) = <max(T(\Delta k))-min(T(\Delta
  k))>$, where $\# (\Delta k) \propto (\Delta k)^{-D}$ for a fractal curve of
  dimension $D$.}}
\bibitem[{not({\natexlab{b}})}]{note_accmodes}
\bibinfo{note}{{The range of the $K$ value is close to the first fundamental
  accelerator mode regime which terminates at $K=7.45$. However, trajectories
  in remaining accelerator mode islands will exit the phase space quickly and
  do not affect the ensuing statistics.}}
\bibitem[{not({\natexlab{d}})}]{note_entryset}
\bibinfo{note}{{We showed the argument for the exitset and not for the entryset
  for the sake of clarity. A corresponding relation for the algebraic
  distribution of lobe widths in the entryset can be derived easily by studying
  the time-reversed map, that again is a chaotic map with the same
  properties.}}
\end{thebibliography}

\end{document}